\documentclass[letterpaper]{article} % DO NOT CHANGE THIS
\usepackage{arabtex}
\usepackage{utf8}
\usepackage{xcolor}
\usepackage{array,booktabs,ragged2e}
\newcolumntype{R}[1]{>{\RaggedLeft\arraybackslash}p{#1}}
\usepackage{aaai25}  % DO NOT CHANGE THIS
\usepackage{times}  % DO NOT CHANGE THIS
\usepackage{helvet}  % DO NOT CHANGE THIS
\usepackage{courier}  % DO NOT CHANGE THIS
\usepackage[hyphens]{url}  % DO NOT CHANGE THIS
\usepackage{graphicx} % DO NOT CHANGE THIS
\usepackage{natbib}  % DO NOT CHANGE THIS AND DO NOT ADD ANY OPTIONS TO IT
\usepackage{caption} % DO NOT CHANGE THIS AND DO NOT ADD ANY OPTIONS TO IT
\usepackage{algorithm}
\usepackage{algorithmic}
\usepackage{newfloat}
\usepackage{listings}
\DeclareCaptionStyle{ruled}{labelfont=normalfont,labelsep=colon,strut=off} % DO NOT CHANGE THIS
\floatstyle{ruled}
\newfloat{listing}{tb}{lst}{}
\floatname{listing}{Listing}
\title{Fabricating Holiness: Characterizing Religious Misinformation Circulators on Arabic Social Media}
\author{
Mahmoud Fawzi, Björn Ross, Walid Magdy}
\affiliations{
    School of Informatics, The University of Edinburgh, Edinburgh, UK
    m.f.g.ibrahim@sms.ed.ac.uk, b.ross@ed.ac.uk, wmagdy@inf.ed.ac.uk
}

\usepackage{bibentry}
\newcommand{\change}[2]{\textcolor{black}{#1}\iffalse{#2}\fi}

\begin{document}

\maketitle

\begin{abstract}
Misinformation is a growing concern in a decade involving critical global events. While social media regulation is mainly dedicated towards the detection and prevention of fake news and political misinformation, there is limited research about religious misinformation which has only been addressed through qualitative approaches.
In this work, we study the spread of fabricated quotes (Hadith) that are claimed to belong to Prophet Muhammad (the prophet of Islam) as a case study demonstrating one of the most common religious misinformation forms on Arabic social media.
We attempt through quantitative methods to understand the characteristics of social media users who interact with fabricated Hadith. We spotted users who frequently circulate fabricated Hadith and others who frequently debunk it to understand the main differences between the two groups. We used Logistic Regression to automatically predict their behaviors and analyzed its weights to gain insights about the characteristics and interests of each group. We find that both fabricated Hadith circulators and debunkers  have generally a lot of ties to religious accounts. However, circulators are identified by many accounts that follow the Shia branch of Islam, \change{Sunni}{#change1.1} Islamic
public figures from the gulf countries, and many \change{Sunni}{#change1.1} non-professional pages posting Islamic content. On the other
hand, debunkers are identified by following academic Islamic scholars from multiple countries and by having more intellectual
non-religious interests like charity, politics, and activism.
\end{abstract}


\section{Introduction}

A hadith is any speech, discussion, action, approval, and physical or moral description attributed to the Prophet Muhammad, whether supposedly or truly \cite{saloot}. It is one of the two major sources of the Islamic religion along with the Quran, which is the holy book of Muslims \cite{khan}.
Historically, people starting fabricating hadiths for political gains as early as 25 years after the death of the Prophet Muhammad  \cite{usman2018fabricated}. Nowadays, social media is facilitating the spread of fabricated hadiths among other types of misinformation \cite{hakak2022digital}. This can be critical because Hadith, as a highly respected resource of knowledge for Muslims, is usually used for argumentation \cite{boutz2017quoting}. This led to serious consequences in the past decade when ISIS used some non-authentic hadiths along with authentic ones as a tool for recruiting members and promoting their agenda worldwide \cite{boutz2019exploiting}.

Many taxonomies, that attempt to break down the topic of misinformation, link it mainly with fake news and political false information \cite{zannettou2019web, molina2021fake, kapantai2021systematic}. However, the volatile global events that occurred during the current decade motivated investigating other specific flavors of misinformation, such as medical misinformation \cite{teplinsky2022online}, corporate misinformation \cite{zhou2024characterizing} and religious misinformation \cite{alimardani2020covid19}. \citet{southwell2022defining} highlighted the difference between political misinformation and scientific misinformation and defined the later as publicly available information that is misleading or deceptive relative to the best available scientific evidence. As for the religion of Islam, Islamic studies have many branches each having different complex methodological approaches for verifying information \cite{bennett2015bloomsbury}. Hadith criticism, which includes determining the authenticity of a religious excerpt claimed to be a hadith, is a crucial branch of Islamic studies \cite{azami2002studies}.

\begin{figure*}
    \centering
    \includegraphics[width=0.99\linewidth]{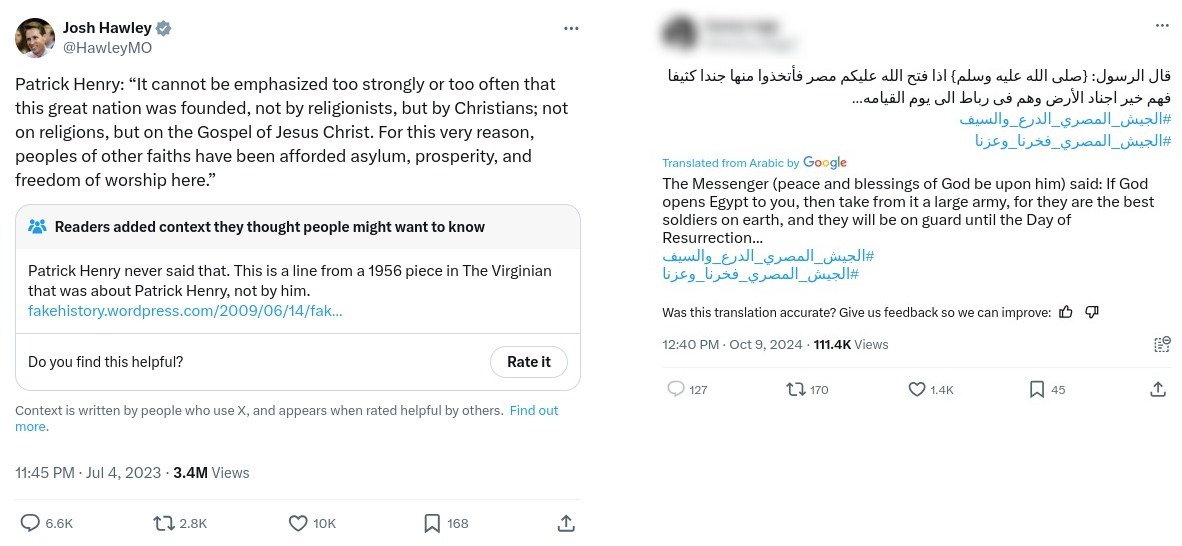}
    \caption{Two tweets with religious misinformation in English (left) and Arabic (right). English tweet received community notes to clarify the context of English misinformation, while the Arabic tweet sharing a non-authentic hadith received nothing.}
    \label{fig:religious_misinformation}
\end{figure*}

There has been very limited work on the presence of Hadith on social media. One recent study by \citet{fawzi2025hadith} shows that fabricated hadiths are widely spread on Arabic social media. Nevertheless, we are not aware of studies that analyzed such special type of misinformation (religious), existing within an understudied community (Muslim world), and spread in an understudied language (Arabic). Although there are a lot of early and recent works that propose techniques for Hadith authentication \cite{ghazizadeh2008fuzzy, gaanoun}, they do not consider the practical contexts and the dynamic formats in which Hadith is exchanged among individuals on the internet. 

\change{Hadith fabrication as a form of}{#change2.5} religious misinformation has also received very little attention \change{on}{#change2.5} social media \change{platforms compared to other religious incidences linked with US politics}{#change2.5}. Figure \ref{fig:religious_misinformation} illustrates this: it shows community notes \change{added by readers}{#change2.5} on X in response to a tweet by an American senator that linked the foundation of the United States to Christianity. In contrast, \change{we are not aware of any tweets including a non-authentic hadith that received a similar interaction to the example shown in Figure \ref{fig:religious_misinformation}}{#change2.5}.
%as it is an even more understudied subset of the topic that involves understudied languages like Arabic and the understudied community of Muslims.
In this study, we try to fill this gap by looking at Hadith fabrication through a social lens. We track the interactions of social media users with fabricated Hadith aiming to check if there are any significant differences between users who circulate fabricated Hadith and those who debunk it. In particular, we address the following two research questions:
\begin{itemize}
    \item{\textbf{RQ1}}: How to identify a user's general behavior towards fabricated Hadith in terms of circulation and debunking?
    \item{\textbf{RQ2}}: What are the key network ties of users that help identify such behaviors?
\end{itemize}

Our analysis provides a better understanding of the nature of users sharing fabricated Hadith in contrast to those who carefully examine its authenticity. This analysis provides a decent building block for systems combating and raising awareness against religious misinformation using the characteristics of users interacting with it. We observe that the networks of both classes of users who interact with fabricated Hadith contain a lot of religious accounts. Circulators are mainly identified by accounts that follow the Shia branch of Islam, Sunni scholars from the gulf countries, and many non-professional \change{Sunni}{#change1.1} pages posting Islamic content. On the other hand, debunkers are identified by \change{Islamic}{#change1.1} scholars from more diverse countries and by having more diverse interests like politics, charity and activism.

\section{Background and Related Work}

%In this section, we start by showing the wide presence of religion in the digital world that motivated our study. Afterwards, we shortly introduce Hadith and some of the quantitative studies that processed it for different purposes with a focus on authentication. Finally, we present some efforts devoted for understanding misinformation and the users involved in it on social media.

\subsection{Hadith}
The word Hadith \setcode{utf8} (in Arabic \<حديث>) literally means communication, story, or conversation: religious or secular, historical or recent \cite{azmi2019computational}. It is used as a proper noun in Arabic to refer to the sayings or actions of Muhammad, the prophet of Islam \cite{binbeshr2021systematic}.
Hadiths are highly regarded in mainstream Islamic faith and play a vital role in shaping Islamic laws.
Each hadith includes a chain of narrators tracing its transmission back to the original source who is Prophet Muhammad. Islamic scholars compiled these narrations into distinct collections one to two centuries after Muhammad’s death.

Islamic scholars have long been aware of the importance of Hadith authentication and have developed an entire science for it. \change{The two main branches of Islam, Sunni and Shia, disagree which sources are credible for verifying Hadith \cite{saloot}. The authenticity of a hadith is based on its chain of narrators who narrated the hadith of the prophet Muhammad, called \textit{Isnad} (in Arabic \<إسناد>). While Sunni scholars trust the companions of the prophet Muhammad as credible sources of Hadith transfer, Shia scholars discredit many of the prophet's companions except for a handful. They only accept hadiths that have been reported by or attributed to the descendants of Muhammad and their supporters \cite{azmi2019computational}.}{#change3.1}

The levels of Hadith authenticity \change{according to Islamic scholars}{#change3.2} are \textit{authentic}, \textit{good}, \textit{weak}, and \textit{fabricated} \cite{abdelaal}. An \textit{authentic} hadith is the one with a strong evidence of belonging to the prophet. A \textit{good} hadith is the one with a less strong yet acceptable evidence of belonging to the prophet. A \textit{weak} hadith has no acceptable evidence of belonging to the prophet. Finally, a \textit{fabricated} hadith has \change{a}{#change0}  strong evidence of not belonging to the prophet. The precise definition of evidence varies among the schools of Islamic scholars but it is generally related to the evaluation of the narrator chain part of that hadith.
There is a disagreement between different Muslim scholars on the authenticity levels of some hadiths, and this discrepancy increases among Sunni and Shia Muslims.
%There are different levels of authenticity of Hadith according to the confidence in the narration chain, including "Authentic", which means scholars are highly confident the hadith is said by Muhammad, "good" or "weak", which are hadiths that are likely said by Muhammad, but there is an issue in the narrators chain leading to less confidence, and "fabricated", which are hadiths that are known to be not said by Muhammad.
%Many politicians realize that most Muslims highly respect Hadith and they often include it in their communication to modulate their political messages. This is not limited to Muslim politicians like Erdoğan\footnote{Erdoğan quoting Hadith: \url{tccb.gov.tr/en/news/542/74878/islam-dunyasi-izzetini-guvenini-nerede-kaybettiyse-orada-aramak-ve-bulmak-zorundadir}}, the president of Turkey, but it also applies to non-Muslim politicians like Jacinda Ardern\footnote{Ardern quoting Hadith: \url{dailysabah.com/asia/2019/03/22/new-zealands-ardern-quotes-prophet-muhammads-hadith-in-solidarity-with-christchurch-victims}}, the prime minister of New Zealand.

Hadith is present in the daily lives of most Muslims. \citet{fawzi2025hadith} show that Hadith is being shared on social media with a high volume every day and that there are multiple incidents in which some Hadith deployments went viral on social media causing social discourse. They also show that some of the widely shared hadiths on social media are actually fabricated. Although extensive research has been conducted on misinformation, there is a notable gap in the study of religious misinformation, specifically\change{,}{#change0} the dissemination of fabricated Hadith among Arab and Muslim communities on social media platforms.
%This was the main motive behind our study as we found that there are no earlier studies that examined Hadith fabrication on social media so we decided to explore it on the same platform where it was found which is X (formerly Twitter).

\subsection{Hadith Digital Authentication}
% In the 1940s, the priest, Father Roberto Busa began recording the writings of Thomas Aquinas in a digital format. This was not only the first religious text to be stored in digital form, but also the main motive to start the \textit{Dead Sea Scrolls} project which would be later the foundation of the modern field of Natural Language Processing (NLP) \cite{busa1958index, ibm}. In the 1980s, Muhammad Mustafa Azmi, an Islamic scholar, decided to create the first digital form of Hadith \cite{al1991note}. Later, some scholars acknowledged the value of this work to Arabic NLP and considered Hadith to be the largest resource covering Classical Arabic literature, a subset of Modern Standard Arabic that is being used nowadays for official Arabic communication \cite{azmi2019computational}.

Some researchers attempted to automate tasks associated with Hadith such as topical classification, authentication, and question answering. There are multiple literature reviews summarizing these attempts \cite{saloot, azmi2019computational, binbeshr2021systematic, sulistio} and showing that a big portion of them focused on Hadith authentication in Arabic. The techniques used for this task included rule-based methods \cite{ibrahim2016frameworks, kabir2018development, kabir2019development}, classical machine learning models \cite{hassaine2016authenticity, najiyah2017hadith, abdelaal}, and transformer models \cite{gaanoun}. One key limitation of these studies is that authentication is mainly performed on hadith texts from literature books. The gold labels for all such texts already exist which questions the utility of such systems. The contexts where Hadith authentication is required include incidents were Hadith is quoted in social media threads, news articles, or official talks. 

There are two valuable Hadith datasets that include the hadith along with its level of authenticity: the LK Corpus \cite{altammami} and MAHADDAT \cite{gaanoun}. The LK Corpus includes 34K hadiths in Arabic, their translation in English, and their degree of authenticity. MAHADDAT includes the authentic Hadith from LK corpus and a list of 2.4K fabricated hadiths that were crawled from multiple Islamic websites. %We use both datasets as our reference for hadith authenticity in addition to consulting multiple Sunni and Shia Islamic scholars.

 \subsection{Online Religious Activity}
%Religious engagement online could be found as early as the 1980s and since then, digital technologies and environments have been shaping religious groups and cultures, and vice versa \cite{campbell2021digital}. This area of scholarship didn't only attract psychologists and sociologists but also scholars from the domain information science \cite{rifat2022integrating, wolf2024still}.

Social scientists have a growing interest in understanding how digital technologies impact the religious and social lives of users. This includes Christianity, e.g. `online churches' \cite{hutchings2017creating} and the digital Bible \cite{hutchings2017design}, Judaism \cite{tsuria2020discourse}, and Islam \cite{fakhruroji2021muslims}. 

Analytical studies that address digital religion are limited compared to the number of digital applications that people use in this domain \cite{buie2013spirituality, wolf2024still}. Nevertheless, the research community has investigated the effect of religion on critical topics like sustainability \cite{rifat2020religion}, women's inclusion \cite{ibtasam2019my}, hate speech \cite{kursuncu2019modeling, albadi2022deradicalizing}, online harassment \cite{bhimdiwala2024fighting}, and health \cite{ibrahim2024tracking}. Focused research on designing better digital religious experiences is also common \cite{kim2022social, markum2023designing, wolf2022spirituality, claisse2023keeping}.% and the COVID-19 pandemic made it more relevant \cite{kaptelinin2021understanding, aduragba2023religion}.

There is another set of studies that investigate the role of religion in global events or daily lives by performing quantitative analysis on big social data. For example, \citet{chandra2021virus} presented the \textit{CoronaBias} dataset which highlighted the Islamophobic hate speech that existed amid the COVID-19 pandemic. \citet{abokhodair} analyzed the presence of Quran verses on Twitter and researched into the motives of people who share it. While analytical research is still lacking studies on religion and spirituality, it lacks even more this type of studies that contribute social knowledge to the community rather than proposing experimental designs \cite{wolf2024still}. We hope that our work will be one of those studies that provide additional knowledge for better understanding of online behavior and better understanding of the online social ecosystem.

There are few studies on religious misinformation that mainly addressed the topic through qualitative or narrative approaches. For example, \citet{alimardani2020covid19} discuss how religious misinformation manipulates people to take fake religious remedies to illnesses in general and the case of COVID-19 in particular. Interestingly, they point out to the problem of the spread of False hadiths. Even quantitative methods like \cite{al2024social} address the bias against specific religious groups through fake stories about them but they do not explore the incidences where religion itself is misused or fabricated.

\subsection{Misinformation Personas}
There are few studies that focus on the users who interact with misinformation compared to the ones that try to detect misinformation \cite{fawzi2024pinocchio}. These studies agree that there is a topology describing how misinformation propagates \cite{morales2023geometry, benkler, mosleh, introne, starbird}. This topology involves complex patterns of information exchange. For example, quotes of officials are reported as news even if those quotes themselves are misleading \cite{benkler}. True news are abused to support fake claims by providing inaccurate interpretations for them \cite{introne}. Public figures contribute to performing what is called "Elite Misinformation" \cite{mosleh} but they do it in a way that does not hold them accountable \cite{starbird}. 

The task of identifying fake news spreaders was performed using different sets of features including retweets \cite{caldarelli}, replies \cite{weinzierl}, language-independent textual features \cite{vogel}, and network structure \cite{rath}. However, digging deeper and investigating the common characteristics of those users and others who do opposite behaviors are less common \cite{mu, fawzi2024pinocchio}. Recent analyses suggest that this characterization is really essential since people do not stop following misinformation spreaders over time \cite{ashkinaze2024dynamics}. Although the percentage of content including misinformation on the web is estimated to be very limited \cite{morales2023geometry}, most of this content still doesn't get debunked, which makes the exploration and capitalization of the debunking behavior an essential research direction besides exploring the spreading behavior \cite{miyazaki2023fake}.

\section{Data \& Methods}

In this section, we explain our approach to collect the required data for the study and to identify the characteristics of users interacting with fabricated Hadith as a case study for religious misinformation.

\subsection{User Group Definitions}
We chose Twitter for our study as it is the same medium where earlier studies found fabricated Hadith \cite{fawzi2025hadith}. We recognize two significant groups of users interacting with fabricated Hadith in Arabic on X (formerly Twitter):
\begin{itemize}
  \item \textbf{Circulators}: The accounts that frequently tweet or retweet scriptures identified as fabricated Hadith.
  \item \textbf{Debunkers}: The accounts that have replied to threads including fabricated Hadith stating they are fabricated or have tweeted warnings about non-authentic hadiths.% with disclaimers and warnings that they are not authentic.
\end{itemize}

While most similar studies focused mainly on circulators, a proper definition of an opposing behavior allows statistical models to capture signals about worrying features that characterize such accounts as well as healthy features that indicate a relatively immune account when found among the network. This definition handles the limitation of earlier studies that consider sharing true news \cite{leonardi} or not sharing some specific fake news \cite{vogel} to be opposite behaviors. It also contributes to the  limited yet important research conducted to characterize debunkers of misinformation \cite{miyazaki2023fake}.

\subsection{Data Collection}
We use Twitter API for all of the stages of the data collection process. The data was collected from March 2023 until the restriction of the API in May 2023. Figure \ref{fig:data_extraction} in the appendix shows visually the detailed steps of data collection that we present below.

\begin{table}[]
    \centering
    \tabcolsep=0.07cm \tiny \begin{tabular}{c c | c c}
        \textbf{Refuting Term} & \textbf{English Translation}  & \textbf{Refuting Term} & \textbf{English Translation} \\ \hline
         \<حديث موضوع>& Fabricated hadith& \<حديث مفبرك>&Fabricated hadith \\
         \<حديث مفترى>& Forged hadith& \<حديث غير صحيح>&Incorrect hadith\\
         \<حديث مكذوب>& Fake hadith & \<حديث كذب على رسول الله>&A lie about the prophet\\
         \<حديث لا يصح>& Incorrect hadith & \<حديث لا أصل له>&hadith with no origin\\
         \<الدرجة: لا يصح>&Degree: Incorrect & \<حديث ضعيف>& Weak hadith \\
         \<الدرجة: موضوع>&Degree: Fabricated & \<حديث ليس صحيح>&hadith not correct\\
         \<حديث لم يرد>&a hadith never mentioned & \<حديث مختلق>&Feigned hadith\\
         \hline
    \end{tabular}
    \caption{A list of the refuting terms in Arabic used for data collection and refutes counting along with their English translation.}
    \label{tab:refuting_terms}
\end{table}
 
\subsubsection{Stage 1:} We collected tweets that include fabricated Hadith and those that include refutes or warnings against fabricated Hadith. To get a list of fabricated hadiths, we used the following three seeds:

\begin{itemize}
    \item \textbf{Seed 1}: The \textbf{MAHADDAT} dataset \cite{gaanoun} includes 2,452 fabricated hadiths scraped from multiple Islamic websites. We used them to query Twitter and 269 of them returned non-empty responses. These responses had a total of 25,809 tweets with 20,284 unique authors.
    \item \textbf{Seed 2}: \textbf{Islamweb}\footnote{\url{islamweb.net}} is the most visited Islamic website worldwide and one of the top five most visited websites under the category of \textit{Faith \& Belief} according to Similarweb\footnote{\url{similarweb.com/top-websites/community-and-society/faith-and-beliefs/}}. Muslims visit it to ask about the rules of Islam or about the perspective of Islam regarding some life situations. The website has a dedicated section for questions about whether a hadith is authentic or not where non-authentic hadiths are grouped together in an archive. We scraped the pages of this archive and use its hadiths to query Twitter in the same way we did with MAHADDAT. 84 out of the 1,113 queries returned non-empty responses resulting in 9,360 hadith tweets with 7,186 unique authors. \change{Since Islamweb was not among the websites used to construct MAHADDAT, we investigated the number of hadiths that exist in both sources. We found that there are only 39 repeated hadiths that exist in both sources which confirms that Islamweb enriched our dataset with a different set of fabricated hadiths.}{#change3.4}
    \item \textbf{Seed 3}: Finally, we used a list of hadith refuting terms to search Twitter. Table \ref{tab:refuting_terms} includes \textbf{14 phrases} that are used by Muslim Arabs to indicate that a hadith is not authentic. We used those phrases to query Twitter, and whenever any of the retrieved tweets are replies, the parent tweet was also collected. This resulted in a total of 3,928 tweets with 3,234 unique authors.
\end{itemize}

%In total, we have collected a set of around 39K tweets with that contain a fabricated Hadith (or a response to a fabricated hadith), tweeted by around 30K unique users.

Since the websites used by MAHADDAT as well as the website Islamweb tend to use the Sunni approach of evaluating Hadith authenticity, we asked \change{a}{#change3.6} Shia scholar to label the top \change{62}{#change3.6} hadiths \change{that}{#change3.6} appeared in our dataset \change{more than once}{#change3.6} to validate if they are also considered fabricated according to the Shia's school. The scholar spotted 15 hadiths that are considered acceptable by Shia while seen fabricated by Sunnis\footnote{\change{The Shia scholar provided us with the four main measures he used to qualify a hadith to be acceptable from a Shia perspective: 1) has an \textit{Isnad}. 2) exists in one of the canonical Shia books like \textit{Al-Kafi}; 3) does not contradict the Quran. 4) does not contradict the fundamentals of the Shia doctrine.}{#change2.1}}. We \change{filtered out}{#change0} the data extracted based on these 15 hadiths so that our study focuses on hadiths that are considered fabricated by both schools.
% \change{The Shia scholar provided us with the four main measures that qualify a hadith to be authentic from a Shia perspective:
% \begin{enumerate}
%     \item The hadith should have an \textit{Isnad}.
%     \item The hadith should exist in one of the canonical Shia books like \textit{Al-Kafi}.
%     \item The hadith should not contradict the Quran.
%     \item The hadith should not contradict  the fundamentals of the Shia doctrine.
% \end{enumerate}
% }{#change2.1}

\subsubsection{Stage 2:} In this stage, we attempt to collect accounts with an opposing behavior to the authors collected in the previous stage. For this we utilize the replies on the collected original tweets and the original tweets of the collected replies.

\begin{itemize}
    \item For seed 1 (MAHADDAT), we collect the 118 replies that included refuting terms and authored by 91 unique users.
    \item For seed 2 (Islamweb), we collect 86 replies that included refuting terms and authored by 75 unique users.
    \item For seed 3, 1,974 refuting tweets were replies to original tweets, we identify the authors of 1,036 unique original tweets.
\end{itemize}

\subsubsection{Stage 3:} In this stage, we collect network interactions of all the authors of original tweets and replies from the three seeds. This includes all the accounts they follow, all the accounts that author what they retweet in their recent timeline, and all the accounts that author what they like in their recent timeline.
The maximum number of tweets that could be retrieved from a user's posting timeline and likes timeline through Twitter API is 3,200 tweets for each. However, there are accounts that have less than 3,200 tweets in their entire timeline. Hence, we have an average of 2,720 posting tweets and 2,040 liked tweets per account.

\subsection{Accounts Shortlisting}
To pick the subset of the accounts to use for our study, we had to find the count of fabricated Hadiths in their recent timeline as well as the number of fabricated hadith refutes. In this way we ensure that we filter out any accounts that have been retrieved due to wrong matching by Twitter API's underlying algorithm. %We shall use again the MAHADDAT dataset as our reference for fabricated Hadith \cite{gaanoun}.

This counting process however is not trivial for hadiths. As mentioned earlier, it is quite common that Muslims include a hadith in a tweet along with other text. It is also common that they only write a part of a hadith and not an entire hadith. For this stage, we adapted the method used by \citet{fawzi2025hadith} for detecting hadith in a tweet, but in our case, we included the fabricated hadiths list. The method can be summarized as follows.

\begin{enumerate}
    \item \textbf{Preprocessing:}
    We preprocess the dataset as well as the timeline as follows:
    \begin{itemize}
        \item \textbf{Arabic Processing:} This includes removing optional characters in Arabic \cite{darwish2012language} and normalizing some letters that can be typed in multiple ways \cite{darwish2014arabic}.
        \item \textbf{Hadith Processing:} A hadith usually starts with a phrase such as \textit{``The messenger of God said''}, \textit{``I heard the prophet saying''}, and \textit{``Muhammad, peace upon him, said''}. These phrases are used interchangeably by individuals to quote a hadith as they do not alter the meaning of the hadith so they are removed.
    \end{itemize}
    \item \textbf{Hashing and Pairing:} The Minhash of all the tweets in the timelines as well as reference hadiths is computed. They are then used to compute the Jaccard similarity between items from both sets efficiently. \change{We apply the same filtering criteria defined by \citet{fawzi2025hadith}.}{#comment3.9} If the similarity score between the tweet and its highest matched hadith exceeds 0.35 then it is considered a match. We validated a sample of 100 matches to validate that this threshold is precise for our dataset and all the records were true positives. Reference hadiths included \textit{fabricated} hadiths from MAHADDAT but also hadiths with other levels of authenticity from LK Corpus \cite{altammami}. This is because we needed to know the total count of hadiths in each user's timeline to calculate the percentage of fabricated hadith in their timelines.
    \end{enumerate}

\textbf{Note:} We also count the number of refutes in each timeline. If a \textit{fabricated} hadith is matched but the tweet includes a hadith refuting term then this case is counted as a refute and not a fabrication.

Although our collection includes more than 30,000 accounts, only 7,382 either have at least one fabricated hadith or one refute in their recent timeline. This is because Twitter API fetched some accounts based on a tweet they tweeted in the past and not in the last 3,200 tweets that are available for the API. We pick accounts that do the behaviors of circulation or debunking most frequently. Our collection includes 559 accounts that spread fabricated hadith two or more times and represents over 5\% of the total hadith they shared. We included all the accounts that never shared a fabricated hadith and that debunked fabricated hadith \change{at least three times}{#change3.11} for the classification experiment, \change{which led to a set of 343 debunker accounts. To have a balanced dataset among both groups, we added to them a set of 216 accounts who debunked only two fabricated hadiths}{#change3.11}. Thus our final user dataset contains a set of 559 \textit{circulators} and 559 \textit{debunkers}. By definition, none of those accounts can fall in both categories. The objective of balancing is to allow the fitting model to equally identify both classes and mitigate bias towards one of them. Balancing does not make the classification problem a realistic one. In fact, in reality, there is a wide spectrum of users who are neither circulators nor debunkers and there are accounts that are not interested in Hadith or never heard of it. This artificial setup is just for the purpose of analysis. %\change{We also acknowledge that our data collection method can potentially result in false negatives when identifying both circulators and debunkers. Overlooked circulators include the users who circulated fabricated hadiths that are not included in the datasets used as a reference as well as the users who made up hadiths themselves or by prompting LLMs. Overlooked debunkers include the users who argue against the authenticity of a hadith in a conversational or indirect way.}{#change3}

\subsection{Network Predictors}

The network activity of users on social media is proven to be a good predictor of multiple private and social attributes. Target attributes could be the expected engagement in some offline event \cite{hu2015predicting}, geolocation \cite{jurgens2015geolocation}, or the stance towards some social or political cases \cite{darwish2022news, magdy2016isisisnotislam}. We are not aware of any work that tried to deploy this strategy on religious data to classify users who interact with it.

We multi-hot encode the interactions of the accounts which means that the predictors vector has a size equivalent to the count of all possible predictors appearing in the entire dataset. The predictors of a given user has the value 1 for any account that this user interacted with and 0 otherwise. We aggregate the sets of predictors modeling liking, retweeting, and following activities of users. Hence, a user can have 1 as a value for their following tie with a page because they follow it and 0 as a value for their liking tie with the same page because they have never liked any of this page's tweets.  We use cross-validation with 90-10\% train-test split just to make sure that the model can fit the data with high accuracy. Then, we retrain the the logistic regression model\footnote{\parbox[t]{\linewidth}{We also tested SVM and ridge regression with the same setup and got almost the same results.}} with the entire dataset to examine their most significant predictors.

\section{Results}
In this section, we first report some statistics about the data then we analyze the significant predictors found within the fitted logistic regression.

\subsection{Data Statistics}

\begin{table*}[h]
\tiny
\resizebox{\textwidth}{!}{\begin{tabular}{ccp{10cm}}
\hline
\textbf{Count} & \textbf{Users} & \textbf{Top Fabricated Hadith (translated)}\\
\hline
233 & 142 & The Messenger of Allah ordered us to present our children to the love of Ali ibn Abi Talib. \\
247 & 89 & Whoever recites Surah Ad-Dukhan during the night, his previous sins will be forgiven. \\
120 & 89 & He, may God’s prayers and peace be upon him and his family, took the pen from Ali’s hand and handed it to Muawiyah. \\
141 & 88 & He, may God’s prayers and peace be upon him and his family, gave Muawiyah a quince. \\
69 & 65 & Whoever wants to look at Adam in his knowledge, Noah in his understanding, Abraham in his wisdom, Yahya in his asceticism, and Moses in his strength, let him look at Ali. \\
77 & 62 & Loving Ali eats away bad deeds as fire eats away wood. \\
65 & 51 & When asked who are the most ascetic person in the world? The prophet referred to the people of his house. \\
371 & 47 & God said to Moses, whoever recites Ayat Al-Kursi after every prayer will be given the reward of the prophets. \\

342 & 43 & Increase your sending of blessings upon me, indeed, Allah appointed an angel to me next to my grave so whenever an individual from my ummah blessings salam upon me the angel says to me: O Muhammad! Verily, so-and-so sends blessings upon you. \\
51 & 41 & I am the city of knowledge and Ali is its gate, so whoever wants knowledge, let him come to the gate.\\

 \hline
\end{tabular}}
\caption{The English translation of the most shared fabricated hadiths in the timelines of the users in our dataset along with the number of unique users sharing them and their total number of occurrences.}
\label{tab:most_popular_english}
\end{table*}

Table \ref{tab:most_popular_english} gives context about the nature of the fabricated hadiths that are shared by the highest number of circulators in our final dataset. The original Arabic version of the hadiths can be found in table \ref{tab:most_popular_arabic} in the appendix.
We consult a professional Sunni Islamic scholar and a professional Shia Islamic scholar to provide us with more context about these hadiths. 
The 1st, 5th, 6th, and 10th hadiths are dedicated to praising Ali, the cousin of the prophet Muhammad, who is a central character in the Shia branch of Islam. The 7th hadith is also dedicated to praise Muhammad's family, which is a central concept for Shia as it supports the claim that Muhammad's family are the only legitimate political successors to him \cite{vaezi2004shia}. The 3rd and 4th hadiths are dedicated to praising Muawiyah, who became a political enemy of Ali after decades of the death of Muhammad. The 3rd hadith reports that Muhammad once took a pen from Ali and handed it to Muawiyah. The 4th hadith reports that Muhammad once gave Muawiyah a quince. Both hadiths symbolize the worthiness of political leadership. These hadiths among others highlight the deployment of fabricated Hadiths in the Sunni-Shia sectarian conflict \cite{mcmahon}. The second common theme that can be detected among these hadiths is claiming that some religious practices lead to God's forgiveness for one's previous sins and to other spiritual rewards. These practices include blessing Muhammad and reading specific verses of the Quran.

\begin{figure}
    \centering
    \includegraphics[width=1\linewidth]{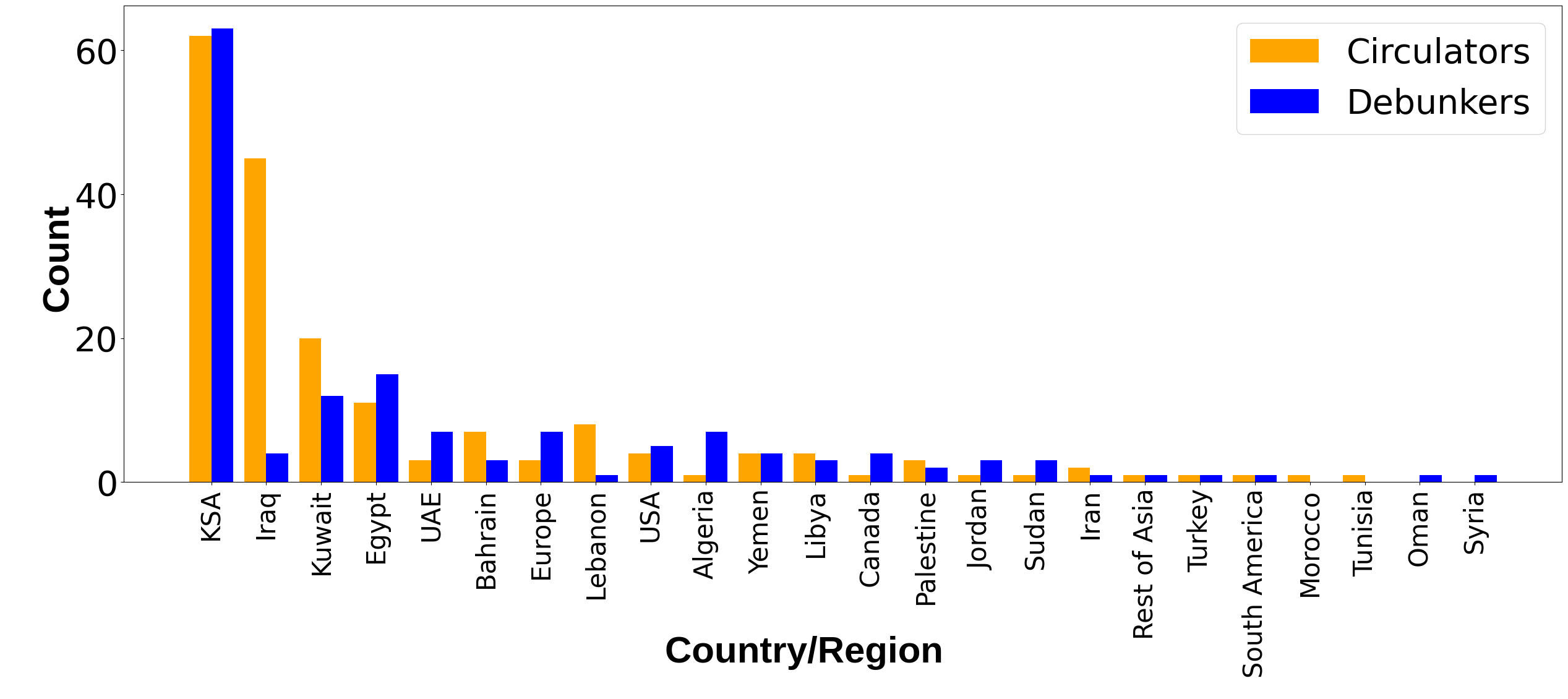}
    \caption{The distribution of user's locations in our dataset as stated in their profiles. 70\% of the users have blank/invalid locations.}
    \label{fig:locations}
\end{figure}

To get more context about the user accounts in our dataset, we inspected the geographical distribution of users \change{manually for accurate reporting of their locations}{#change3.x}. Figure \ref{fig:locations} shows that the users are distributed over a diverse set of Arab and non-Arab countries. Saudi Arabia has the largest portion of both circulators and debunkers. Most countries where circulators outnumber debunkers have a considerable population of Shia Muslims like Iraq, Kuwait, Lebanon, and Bahrain. There is still a considerable number of circulators in countries where \change{Sunni Muslims are a majority and }{#change1.1} Shia Muslims are a small minority like Egypt, UAE, and Saudi Arabia. In all western communities debunkers outnumber circulators.

\begin{figure}
    \centering
    \includegraphics[width=1\linewidth]{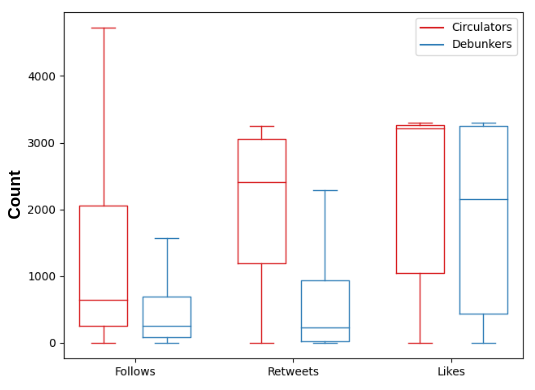}
    \caption{The average number of overall interactions found for each class.}
    \label{fig:boxplots}
\end{figure}

We also checked the distributions of different types of interactions and connections of each group, see Figure \ref{fig:boxplots}. In general, circulators tend to have a \change{significantly\footnote{\change{according to Welch's significance t-test with p-value of 0.0001 for distributions of retweets, follows, and likes.}{#change2.2}}}{#change2.2} higher number of interactions than debunkers. It is also worth noting that the percentage of retweets in the timelines of circulators is much higher (75.8\%) than debunkers (27.9\%). This means that they have higher tendency to share content authored by others compared to debunkers.

% \begin{table*}[]
% \begin{center}
% \small
% \begin{tabular}{lcccccccc} 
%  \hline
%  Users & \multicolumn{2}{c}{Avg Follows} & \multicolumn{2}{c}{Avg Likes} & \multicolumn{2}{c}{Avg Retweets} & Avg \% retweets \\
%  Group & overall & intraclass & overall & intraclass & overall & intraclass & in timeline \\
%   \hline
% Circulators & 2001 & 8.4 & 2172 & 5.0 & 2307 & 7.5 & 75.8\% \\
% Debunkers & 804 & 1.8 & 1907 & 4.6 & 669 & 1.6 & 27.9\% \\
%  \hline
% \end{tabular}
% \end{center}
%     \caption{The average number of overall interactions for each users group and the average number of intraclass interactions. Average likes and retweets are calculated over the 3,200 post threads and the 3,200 like threads available for each account only and not their entire timelines}
%     \label{tab:diversity}
% \end{table*}

% We also checked how far the users in our dataset are interconnected as the significance of our analysis would considerably be reduced if the users belong to a closed group where they interact with each other. Table \ref{tab:diversity} shows that the intraclass connections of different types among both user groups are very few compared to the number of overall connections. This supports the impression about diversity concluded after the inspection of the geographical distribution. It is also worth noting that the percentage of retweets in the timelines of circulators is much higher than debunkers. This means that they have higher tendency to share content authored by others compared to debunkers.

\begin{table}[]
\begin{center}
\begin{tabular}{l c c c c} 
 \hline
  & Accuracy & Precision & Recall & F1-Score \\
 \hline
 Circulators & - & 92.2\% & 82.3\% & 87\% \\ 
 Debunkers & - & 84.1\% & 93.0\% & 88.3\% \\ 
 Macro & 87.7\% & 88.2\% & 87.7\% & 87.6\% \\
 \hline
\end{tabular}
\end{center}
    \caption{Precision, recall, and F1-score of the logistic regression model whose predictors get analyzed further. Accuracy and macro values are reported.}
    \label{tab:regression-metrics}
\end{table}

\subsection{Analysis}

Logistic regression is able to distinguish fabricated Hadith circulators from debunkers with an accuracy of 87.7\% \change{as shown in Table \ref{tab:regression-metrics}}{#change3.15}. The high accuracy obtained by the fitted model qualify its most significant predictors to be identifiers for both classes. The top 100 predictors with highest positive and negative coefficients assigned by the logistic regression are inspected further where positive ones identify the circulators class and negative ones identify debunkers.

\begin{figure}
  \centering
  \includegraphics[width=\linewidth]{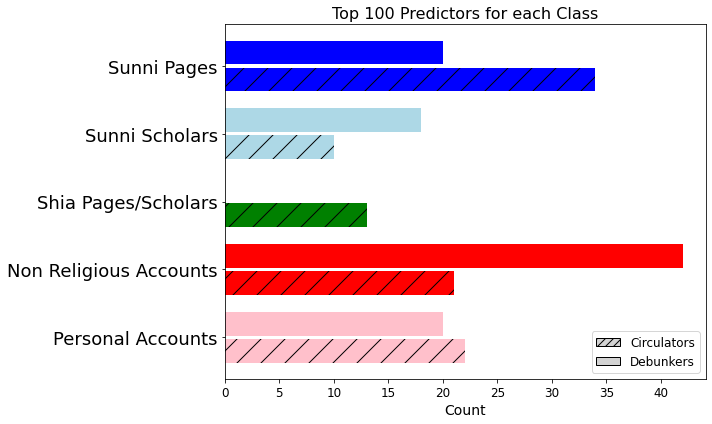}
  \caption{The distribution of the top 100 significant predictors identifying Circulators and Debunkers in the logistic regression.}
  \label{fig:predictors}
\end{figure}

We classify the predictors into five categories: Sunni Pages, Sunni Scholars, Shia Pages/Scholars, Non-religious Accounts, and Personal accounts as shown in Figure \ref{fig:predictors}. We notice that a big portion of the most significant features consists of accounts related to religion. 
%For some topics like climate change and legalization of abortion, most of the significant features identifying users' behaviors towards these topics weren't directly related to them \cite{aldayel}. 
There are significant differences between circulators and debunkers in terms of the portion of each category in their top 100 predictors and the nature of the accounts in each portion. Below, we go through each category in detail:

\textbf{Sunni Pages:} This category represents religious accounts following the Sunni branch of Islam sharing Islamic content. There are more accounts from this category identifying circulators compared to debunkers. By inspecting these accounts further, we find that the ones on the circulators side are usually not related to a professional website or institution. Some of them just identify themselves as a group of social media influencers posting religious content. In case of debunkers, those accounts usually represent professional websites that are moderated by professional Islamic scholars (e.g \texttt{saaidnet1}\url{saaid.org}) and professional Islamic institutions (e.g.   \texttt{TPCV\_SSA} and \texttt{olamayemen}).

\textbf{Sunni Scholars:} For circulators, all of the accounts under this category belong to gulf countries. On the other hand, for debunkers, there are scholars from other areas like Ali al-Qaradaghi (\texttt{Ali\_AlQaradaghi}) from Turkey and Muhammad al-Dido (\texttt{ShaikhDadow}) from Mauritania. Additionally, we spotted \texttt{ashrafgharib11} who defines himself as an anti-shia researcher in the Shia religion and \texttt{AntiShubohat} who defines himself as a Muslim researcher against heresies related to Islam.

\textbf{Shia Pages/Scholars:} This category only exists for circulators, but not for debunkers, as shown in Figure \ref{fig:predictors} and its accounts have the highest coefficients among all the most significant features. However, being associated with circulators does not necessarily imply that they circulate fabricated hadith themselves. It just means that they are distinctive predictors in the network of the circulators. \change{Nevertheless, it is important to note in this context that Sunni pages/scholars associated with circulators as features are more than twice the Shia features. The presence of many religious features from both Sunni and Shia branches of Islam confirms that the behavior of Hadith fabrication is not solely linked to one of them but appears within both of them.}{#change1.1}

\textbf{Non-religious Accounts:} Features from this category on the debunkers side are twice more than the circulators side. This shows that debunkers are more identifiable through their non-religious interactions.  Interestingly, for circulators, accounts from this category include accounts that post pornographic content, and others that are anti-Islam and pro-Israel (e.g. \texttt{BellaZamoria}). Additionally, we also found many accounts related to movies or shows (e.g. \texttt{NetflixMENA} and \texttt{AflamWorld}). We noticed particular interest in Korean shows through accounts (e.g. \texttt{jjklve} and \texttt{bts\_nanak}). Finally, we found that ties with \texttt{NASA}, \texttt{Google}, \texttt{SpaceX}, \texttt{Tesla}, and \texttt{BillGates} identify circulators. These shows that circulators \change{a}{#change0} have wide range of interest\change{s reflecting that}{#change0} they do not \change{necessarily have to }{#change0} be religious, which might explain being less careful when sharing Islamic content that happens to be fabricated sometimes.

For debunkers, accounts include news sources such as \texttt{cnnarabic} and \texttt{SaudiNews50}, journalists (e.g. \texttt{Omar\_Madaniah}), and the account of \texttt{elonmusk}. Furthermore, we recognize charities (e.g \texttt{insan\_kwt} and \texttt{Tarahumkw}) in addition to accounts promoting activism such as \texttt{turkistantuzbah} which is dedicated for the causes of Islamic minorities worldwide, \texttt{hureyaksa} which is against authoritarian regimes in the Middle East, and \texttt{Ali\_Albukhaiti}, an activist against the Houthi movement in Yemen, Hezbollah, and Iran which are entities that constitute a political Shia \change{axis}{#change0} in the Middle East.

\textbf{Personal Accounts:} There are two clusters of personal accounts that identify circulators. The first includes eight accounts that identify themselves as followers of the Shia branch of Islam and the second one includes five accounts that come from Indonesia. Indonesia's first language is not Arabic which makes checking the authenticity of Hadith for Indonesian Muslims a harder task than Arabs. For debunkers, we found two accounts from this category identifying themselves as heresy fighters.

%Table \ref{tab:features-circulators-debunkers} shows samples of the list of the accounts found in the top 100 predictors for each group.
Out of the 100 most predictors for circulators, there are two accounts from our seed list of circulators, namely \texttt{alsalaf\_1}, a Sunni page and \texttt{AhlulbaytSays} a Shia page. This might indicate that these pages have some influence in spreading fabricated hadiths. On the other hand, five accounts belong to our debunkers seed list are in the top 100 significant predictors of debunkers. Those include one personal account,  two professional Islamic pages (\texttt{ConsHadith} and \texttt{alsunna\_way}), and two Islamic scholars (\texttt{almonajjid} and \texttt{hadeeth\_tarifi}), which shows their influence in debunking fabricated hadith online.

\section{Discussion}

To the best of our knowledge, this is the first study to address the topic of religious misinformation by trying to understand the characteristics and the interests of the users involved in it. Our study highlights and quantitatively frames a form of the understudied topic of religious misinformation. Earlier research only addressed religious misinformation by either qualitatively examining and linking some incidents then drawing conclusions based on them \cite{alimardani2020covid19} or measuring quantitatively the reactions of religious communities towards a specific topic \cite{lee2022covid}. However, no research is attempting to dig deeper into the nature of fabricated religious texts that circulate and potentially cause these incidents or reactions. In addition, no research attempts to understand the characteristics of users who interact with such type of misinformation. We questioned in RQ1 the possibility of classifying fabricated Hadith circulators against debunkers using the network ties of these users. The results we obtained showed that both classes are highly separable through their network ties by a simple logistic regression model. 

This easy distinction between both classes motivated our RQ2 where we examine the key features in the networks of each group of users that identify them. We found that both classes are generally identified by their interactions with a lot of religious accounts where the nature of these accounts differ for each group. Religious accounts that identify circulators include Shia pages and scholars, \change{Sunni}{#change1.1} non-professional pages sharing religious content, and \change{Sunni}{#change1.1} professional Islamic scholars from gulf countries. Religious accounts identifying debunkers include pages that are linked with professional websites in addition to Islamic scholars who are from a diverse set of countries. 

Debunkers are more identifiable through their non-religious interactions that include ties with news sources, journalists, charities, and activists. \change{This finding can potentially contribute to the ongoing research debate about the relationship between misinformation and activism . While some research works claim that online activism is associated with misinformation \cite{sinpeng2021grassroots}, a recent study claims finding little evidence to support that  \cite{jalli2025viral}. Our study detects activism within the networks of debunkers which motivates doing further research on Arabic speakers to validate this effect.}{#change1.1} The non-religious interactions identifying circulators included pornographic and atheistic accounts which indicates that posting religious content does not necessarily indicate religiosity.

The limited number of personal accounts among the features compared to public and famous ones suggests that the propagation of fabricated Hadiths takes the form of elite misinformation described by \citet{mosleh} more than the form of horizontal misinformation described by \citet{vladutescu}. In elite vertical misinformation, public figures act by influence on a passive crowd who are in an inferior position. While doing this, these influential accounts sometimes take precautions to distance themselves from being held accountable for disinformation \cite{starbird}. In horizontal misinformation, crowds are more active where an individual is usually a source and a target for misinformation simultaneously.

\subsection{Implications}

%This understanding is essential for the HCI community and for any initiatives or interactive designs that are willing to limit the negative impact of misinformation. HCI research already acknowledges the lack of this kind of studies that attempt to understand the users and help designers implement guided designs \cite{wolf2024still}.
Our findings raise questions about the ethical privacy concerns on social media especially those related to religious beliefs. We showed that the social interactions of users can easily be used to reveal their religious awareness and their ability to distinguish authentic religious texts from fabricated ones, a property that many individuals would not be comfortable to share. While there are many incidents where religion is used to intensify conflicts, our findings reveal \change{another}{#change3.18} threat that arises from the possibility of doing religious provocation on a targeted individual basis. Social media platforms are encouraged to develop techniques that measure the possibility that users get manipulated using religion and warn users who are found to be in this position. \change{The platforms should also raise awareness about the presence of this type of misinformation by reporting analytics about the topic in a similar way to the project of Twitter Transparency Report \cite{twitter_transparency}. It would also be helpful to add religious misinformation as a possible explicit reason for reporting social media threads as spam/harmful.}{#change2.6}

A lot of social studies highlight the negative impact of religion on people's cognitive abilities and claim that atheists and agnostics are more reflective than religious believers \cite{pennycook2016atheists}. They also link being more religious to being more prone to believe fake news \cite{bronstein2019belief}. While this can be true in some communities, it does not necessarily have to be generally true especially within communities where the majority of people are religious. Our study reveals a new dimension of studying religious people where they can be classified into multiple groups that have totally different behaviors towards misinformation. This motivates inspecting the effects of religion that are studied before while considering that religious people can be classified into multiple groups.

Although our study addresses the topic of Hadith fabrication in a neutral way towards Sunni and Shia schools of Hadith evaluation, it highlights that the available resources to study Hadith from the Sunni perspective are much more than those available for Shia. Studying Hadith from Shia's point of view requires consulting professional Islamic Shia scholars who are not accessible to everyone in the research community.  We believe that there is a need for datasets that evaluate Hadith from the Shia perspective. This will help researchers to study the religious beliefs of Shia communities more smoothly. The research community already reports many difficulties Shia Muslims are experiencing with their religious expression \cite{alshehri2023comfort}

Finally, our study sheds light on Hadith as a text in general and the topic of the spread of fabricated Hadith in particular. This interesting complex phenomenon manifests a lot of observations about the different cultural backgrounds of social media users in the Middle East and the disputes they have in a religious flavor. There are a lot of quantitative works that addressed social discourse in areas of conflict like Ukraine \cite{iulia2024} and they did not find a significant role for religion in such conflicts. However, in understudied parts of the world like the Middle East as well as India and Sri Lanka, religion has a significant role in conflicts \cite{koch2019unmasking}. Our study motivates exploring this effect further through quantitative methods rather than the limited existing qualitative ones.

\subsection{Limitations \& Future Work}
Our analysis to Islamic religious misinformation is limited to Hadith as the religious text. However, other religious texts are also often used for misinformation as we showed in Figure \ref{fig:religious_misinformation}. This includes users sharing quotes by scholars or historic figures that are not true. However, the main challenge studying such type of misinformation is the absence of datasets or algorithms to spot them. It can be an interesting direction to create such resources. This analysis can be applied to similar texts in different religions. 

Secondly, our study only considers the Hadith in Arabic text. However, there are hundreds of millions of Muslims who report Hadith in other languages such as Turkish, Farsi, Urdu, and Malay \cite{sulistio}. Exploring how fabricated Hadith spread among these non-Arabic communities would be an interesting future direction, especially when compared to the findings of this study on the Arab world. \change{Additionally, while our study only investigates disputes about fabricated hadith, it is interesting to characterize users who attempt to debunk Hadith that is actually authentic and whether this should be classified as a form of misinformation or intended disinformation.}{#change3.8}

\change{We acknowledge that our strict data collection methodology can potentially result in missing many circulators and debunkers. Overlooked circulators include the users who circulated fabricated hadiths that are not included in the datasets used as a reference as well as the users who made up hadiths themselves or by prompting LLMs. Overlooked debunkers include the users who argue against the authenticity of a hadith in a conversational or indirect way. Nevertheless, our data collection method still retrieves an unbiased representative sample of both behaviors with a high precision which is more necessary for characterizing such behaviors than recall to avoid potential noise in the data.}{#change1.2}

Another important limitation is that our study only investigates Hadith on Twitter (currently X). Hadith exchanges is also very popular on other platforms such as Telegram and Whatsapp \cite{sauda2020one}. Finally, we believe that setting up surveys and interviews with Muslim users can provide good qualitative insights about the awareness and motives behind circulating or debunking fabricated Hadith. They can also provide more context about the possible consequences of this type of misinformation.

\section{Conclusion}
In this work, we introduced the topic of Hadith fabrication as a form of religious misinformation. We identified tens of thousands of threads on social media that include fabricated Hadith and we collected multiple sets of users who interact with it. We attempted to gain more insights about the differences between users who circulate fabricated Hadith and those who debunk it. Both classes were found to be tied with a lot of religious accounts but the accounts tied to each class have a different nature. Circulators are tied with Shia accounts, non-professional \change{Sunni}{#change1.1} pages, and professional \change{Sunni}{#change1.1}  Islamic scholars from gulf countries while debunkers are tied with more professional pages and professional Islamic scholars from a diverse set of countries. The non-religious interests that identify circulators are limited to entertainment while debunkers are identified by interests related to charity, politics, and activism.

\bibliography{aaai25}

\newpage

% \subsubsection{Ethical Disclaimer}: In our study, we ensure using the term misinformation and not disinformation as we don't claim any assumptions about the intentions of users sharing fabricated Hadith or their prior knowledge about it. This means that our labels for accounts are just meant to frame the observed social behaviors to enable further analysis for their causes. The strict classification of users accounts is not among the objectives of our study. We also refrain from exposing any details that can help reveal the identity of the users whose accounts are being studied. Finally, we obtained an ethical approval from our host institution to conduct this study.

\appendix

\newcommand{\answerYes}[1]{\textcolor{blue}{#1}} 
\newcommand{\answerNo}[1]{\textcolor{teal}{#1}} 
\newcommand{\answerNA}[1]{\textcolor{gray}{#1}} 
\newcommand{\answerTODO}[1]{\textcolor{red}{#1}} 

\section{A Ethics Checklist}

\begin{enumerate}

\item For most authors...
\begin{enumerate}
    \item  Would answering this research question advance science without violating social contracts, such as violating privacy norms, perpetuating unfair profiling, exacerbating the socio-economic divide, or implying disrespect to societies or cultures?
    \answerYes{Yes, In our study, we ensure using the term misinformation and not disinformation as we don't claim any assumptions about the intentions of users sharing fabricated Hadith or their prior knowledge about it. This means that our labels for accounts are just meant to frame the observed social behaviors to enable further analysis for their causes. The strict classification of users accounts is not among the objectives of our study. We also refrain from exposing any details that can help reveal the identity of the users whose accounts are being studied. Finally, we obtained an ethical approval from our host institution to conduct this study.}
  \item Do your main claims in the abstract and introduction accurately reflect the paper's contributions and scope?
    \answerYes{Yes}
   \item Do you clarify how the proposed methodological approach is appropriate for the claims made? 
    \answerYes{Yes}
   \item Do you clarify what are possible artifacts in the data used, given population-specific distributions?
    \answerYes{Yes, in section 3 entitled Data \& Methods}
  \item Did you describe the limitations of your work?
    \answerYes{Yes}
  \item Did you discuss any potential negative societal impacts of your work?
    \answerNA{NA, the study describes characterizes two social behaviors without claiming that misinformation is intentional.}
      \item Did you discuss any potential misuse of your work?
    \answerYes{Yes, we highlight that the easy identification of the religious preferences of users can be misused.}
    \item Did you describe steps taken to prevent or mitigate potential negative outcomes of the research, such as data and model documentation, data anonymization, responsible release, access control, and the reproducibility of findings?
    \answerYes{Yes, and we obtained an ethical approval from our host institution to conduct the study.}
  \item Have you read the ethics review guidelines and ensured that your paper conforms to them?
    \answerYes{Yes}
\end{enumerate}

\item Additionally, if your study involves hypotheses testing...
\begin{enumerate}
  \item Did you clearly state the assumptions underlying all theoretical results?
    \answerNA{NA}
  \item Have you provided justifications for all theoretical results?
    \answerNA{NA}
  \item Did you discuss competing hypotheses or theories that might challenge or complement your theoretical results?
    \answerNA{NA}
  \item Have you considered alternative mechanisms or explanations that might account for the same outcomes observed in your study?
   \answerNA{NA}
  \item Did you address potential biases or limitations in your theoretical framework?
    \answerNA{NA}
  \item Have you related your theoretical results to the existing literature in social science?
    \answerNA{NA}
  \item Did you discuss the implications of your theoretical results for policy, practice, or further research in the social science domain?
    \answerNA{NA}
\end{enumerate}

\item Additionally, if you are including theoretical proofs...
\begin{enumerate}
  \item Did you state the full set of assumptions of all theoretical results?
    \answerNA{NA}
	\item Did you include complete proofs of all theoretical results?
    \answerNA{NA}
\end{enumerate}

\item Additionally, if you ran machine learning experiments...
\begin{enumerate}
  \item Did you include the code, data, and instructions needed to reproduce the main experimental results (either in the supplemental material or as a URL)?
    \answerNA{NA, Only simple statistical models are used}
  \item Did you specify all the training details (e.g., data splits, hyperparameters, how they were chosen)?
    \answerNA{NA, Only simple statistical models are used}
     \item Did you report error bars (e.g., with respect to the random seed after running experiments multiple times)?
    \answerNA{NA, Only simple statistical models are used}
 	\item Did you include the total amount of compute and the type of resources used (e.g., type of GPUs, internal cluster, or cloud provider)?
    \answerNA{NA, Only simple statistical models are used}
     \item Do you justify how the proposed evaluation is sufficient and appropriate to the claims made? 
    \answerNA{NA, Only simple statistical models are used}
     \item Do you discuss what is ``the cost`` of misclassification and fault (in)tolerance?
    \answerNA{NA, Only simple statistical models are used}
  
\end{enumerate}

\item Additionally, if you are using existing assets (e.g., code, data, models) or curating/releasing new assets...
\begin{enumerate}
  \item If your work uses existing assets, did you cite the creators?
    \answerYes{Yes, we cite the datasets and adopted mechanisms}
  \item Did you mention the license of the assets?
    \answerNA{NA}
  \item Did you include any new assets in the supplemental material or as a URL?
    \answerNA{NA}
  \item Did you discuss whether and how consent was obtained from people whose data you're using/curating?
    \answerNA{NA}
  \item Did you discuss whether the data you are using/curating contains personally identifiable information or offensive content?
    \answerNA{NA}
\item If you are curating or releasing new datasets, did you discuss how you intend to make your datasets FAIR (see \citet{fair})?
\answerNA{NA, we do not intend to release a dataset}
\item If you are curating or releasing new datasets, did you create a Datasheet for the Dataset (see \citet{gebru2021datasheets})? 
\answerNA{NA, we do not intend to release a dataset}
\end{enumerate}

\item Additionally, if you used crowdsourcing or conducted research with human subjects...
\begin{enumerate}
  \item Did you include the full text of instructions given to participants and screenshots?
    \answerNA{NA, we neither used crowdsourcing nor conducted research with human subjects.}
  \item Did you describe any potential participant risks, with mentions of Institutional Review Board (IRB) approvals?
    \answerNA{NA, we neither used crowdsourcing nor conducted research with human subjects.}
  \item Did you include the estimated hourly wage paid to participants and the total amount spent on participant compensation?
    \answerNA{NA, we neither used crowdsourcing nor conducted research with human subjects.}
   \item Did you discuss how data is stored, shared, and deidentified?
   \answerNA{NA, we neither used crowdsourcing nor conducted research with human subjects.}
\end{enumerate}

\end{enumerate}

\begin{figure*}
    \centering
    \includegraphics[width=0.8\linewidth]{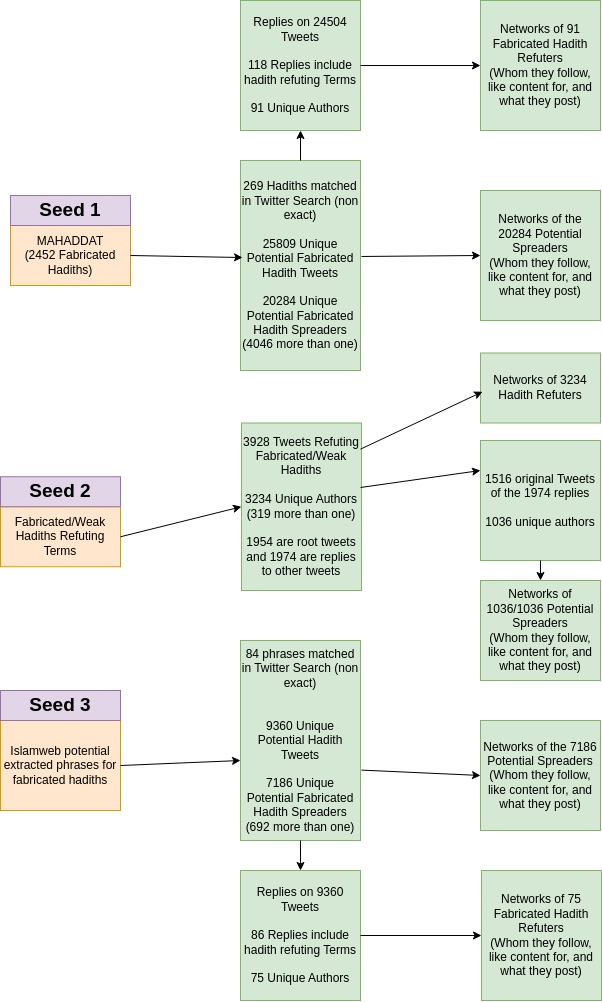}
    \caption{The steps of collecting the dataset. Yellow blocks represent the initial seeds used as a starting point. Green blocks represent data collected using Twitter API that are related to the circulators and debunkers of fabricated Hadith.}
    \label{fig:data_extraction}
\end{figure*}

\begin{table*}
\centering
\begin{tabular}{ p{1cm} p{1cm} R{15cm}}
\toprule
\multicolumn{1}{l}{\textbf{Count}} & \textbf{Users} & \textbf{Top Fabricated Hadith}\hspace{10.5cm}\mbox{ } \\ 
\midrule
233 & 142 & \<أمرنا رسول الله صلى الله عليه وسلم أن نعرض أولادنا على حب علي بن أبى طالب> \\
247 & 89 & \<من قرأ سورة الدخان في ليلة غفر له ما تقدم من ذنبه> \\
120 & 89 & \<أنه صلى الله عليه وآله وسلم أخذ القلم من يد علي فدفعه إلى معاوية
> \\
141 & 88 & \<أنه صلى الله عليه وآله وسلم دفع إلى معاوية سفرجلة> \\
69 & 65 & \hspace{0.85cm}\<من أراد أن ينظر إلى آدم في علمه ونوح في فهمه وإبراهيم في حكمه ويحيى في زهده وموسى في بطشه\\ فلينظر إلى علي> \\
77 & 62 & \<حب علي يأكل السيئات كما تأكل النار الحطب> \\
65 & 51 & \<من أزهد الناس في العالم فقال صلى الله عليه وآله وسلم أهل بيته> \\
371 & 47 & \<قال الله لموسى من قرأ آية الكرسي في دبر كل صلاة أعطيته ثواب الأنبياء> \\

342 & 43 & \hspace{0.95cm}\<أكثروا الصلاة علي فإن الله وكل بي ملكاً عند قبري فإذا صلى علي رجل من أمتي قال لي ذلك الملك يا  \\محمد إن فلان بن فلان صلى عليك الساعة> \\
51 & 41 & \<أنا مدينة العلم وعلي بابها فمن أراد العلم فليأته من بابه>\\
\bottomrule
\end{tabular}
\caption{The most shared fabricated hadiths in the timelines of the users in our dataset along with the number of unique users sharing them and their total number of occurrences.}
\label{tab:most_popular_arabic}
\end{table*}

\end{document}